**ENHANCED OPTICAL, STRUCTURAL AND ANTIBACTERIAL PROPERTIES OF ZnO DOPED $TiO_2$ COMPOSITES**

Mehmet Eymen Sümer[a], Bengü Özuğur Uysal[b]*

[a] *Computational Biology and Genetics, School of Graduate Studies, Kadir Has University, Cibali, Fatih, Istanbul 34083, Turkey*

[b] *Faculty of Engineering and Natural Sciences, Kadir Has University, Cibali, Fatih, Istanbul 34083, Turkey*

** Corresponding author*

*Tel: +90 212 533 65 32 / 1345*

*Fax: +90 212 533 65 15*

*e-mail(s): bozugur@khas.edu.tr*

## ABSTRACT

In daily life, bacterial infections are common issues for human health. Thus, researchers are searching for valuable materials to prevent antibacterial infections and environmental pollutants. Thin film is a well-known application for photocatalytic and biological activity. Hence, thin-film formation is an excellent way to overcome these problems due to its nanosize thickness and enhanced monolayer or a multilayered structure. The most common material to produce a thin film is titanium oxide ($TiO_2$). It has various traits to enhance the activity of thin films, such as structural and optical properties. However, single usage of pure $TiO_2$ has some limitations over these problems. Therefore, new techniques need to be implemented to overcome these limitations, and it is called doping. Doping is a standard method for manipulating material properties to provide enhanced functionality. Thus, ZnO was selected as a dopant due to its good bandgap energy and high electron activity. Therefore, this research mainly focused on the antibacterial, structural, and optical activity differences of pure $TiO_2$ and ZnO doped $TiO_2$. The sol-gel method was used in this research for several different thin film deposition methods due to its easy progression at room temperature, low cost, and homogeneity traits. Antibacterial activity of pure and doped $TiO_2$ thin films was analyzed by the standard of ISO 22196 protocol against gram-positive "*Staphylococcus aureus*" and gram-negative "*Escherichia coli*". As a result, XRD and UV-vis spectrophotometer measurements show that our dopant ZnO efficiently enhances the bandgap energy of pure $TiO_2$, and the correlation between dispersibility and homogeneity was achieved in the concentration range ZTA-B-R (5-10).

## 1. INTRODUCTION

Material science and bioengineering are important majors of human life. Thus, the collaboration of these two majors can be beneficial for biotechnological advancement. In recent years, environmental problems have increased rapidly due to the waste product of many applications[1]. Due to the high amount of waste products, many researchers use nanocomposites (NC) as a bypass mechanism for the antibacterial inhibition of waste products. Thin-film application is a well-known method for enhancing the activity of antibacterial inhibition and photocatalytic activation by using nanocomposites. NCs are critical materials for many biological applications [2] [3]. Therefore, finding proper concentration and combination of nanocomposites in the thin-film application is essential for the activity level of the material. However, in some applications, the single nanoparticle may not be efficient, and some material addition is required to enhance the activity level of the material, which is called a dopant. Therefore, dopant materials are generally used to enhance the activity level of the initial material by changing the materials' initial bandgap energy. Thus, the addition and selection of proper dopant material is crucial for a succession of research because not every dopant can synergize with every nanoparticle. Therefore, before starting this research, most of the dopants and their collaboration with nanoparticles were determined. In the selection period, three main criteria arise; the first one is dopant material suitable for this experimental methodology, the second one is dopant’s ability to enhance antibacterial effects, and the third one is dopant synergize with an initial nanoparticle. Therefore, before starting any antibacterial experiment with the nanoparticle, these topics need to be checked for further implementation. Primary materials need to be multifunctional. Titanium dioxide ($TiO_2$) is a transition metal, and it is a valuable material for material science and biomedicine.

Titanium dioxide ($TiO_2$) has a bright white color, and it has three crystalline phases: anatase, brookite, and rutile. Titanium dioxide ($TiO_2$) is an excellent semiconductor in terms of bandgap energy, such as anatase 3.20 eV, rutile 3.00 eV, and brookite 3.13 eV [4-6]. Titanium oxide ($TiO_2$) is one of the most favorable materials for its multifunctional traits. Titanium oxide ($TiO_2$) is also a promising material due to its photocatalytic, semi-conductive, high degradation, easy to process, and low-cost traits [7]. In addition, titanium dioxide ($TiO_2$) has multifunctional traits such as high recombination rate, biodegradability, and photocatalytic activity in higher temperatures [8]. The anatase and brookite phases naturally transform into the rutile phase under high temperatures, and this trait makes the rutile phase the most stable phase [9]. Its morphology highly depends on its phase differences and chemical manipulation, such as dopant addition. Anatase, brookite and rutile are the three main phases of $TiO_2$ [7]. All these phases have distinct abilities and provide different advantages due to their different structural and chemical activities. For example, anatase has a tetragonal structure, and brookite has an orthorhombic structure [9]. Moreover, the anatase phase is the initial phase formed in many applications, and then it transforms into brookite and rutile in high temperatures, respectively.

Anatase has the largest bandgap than other phases, and this trait makes anatase more photocatalytic than the others [7]. However, rutile can be more stable in higher temperatures and has the highest refractive index [8]. Transformation of anatase to the rutile phase is irreversible. Brookite is rarely seen compared to anatase and rutile because its production is too hard for lab conditions. However, brookite has the most significant cell volume. That feature makes the brookite phase valuable for biomedical and environmental treatments [9]. $TiO_2$ is used in various areas such as dye-sensitized solar cells, ion batteries, gas sensors, filters, light sensors, humidity sensors, hydrogen sensors, tissue engineering, biomedical and biodegradable treatments, industrial dyes, textiles, and environmental remediation [8]. However, sometimes using a single material can produce some problems because of materials' limitations but these limitations are overcome by using dopants. Many studies show that mixed phases are more impactful than pure $TiO_2$ because of their limitations [10]. A cross-section of these limitations needs to be handled correctly to get the desired result [11].

Therefore, the additional dopants -as mentioned before-need to be selected to synergize with titanium oxide and overcome its limitations. Every dopant has different bandgap energy, recombination rate, electron mobility, surface structure, and other features. Thus, dopants generally enhance the activity of primary material due to their beneficial effects on structural, morphological, chemical, and physical traits (Khlyustova et al., 2020). Therefore, dopants' addition may manipulate the $TiO_2$ main features such as surface area, recombination rate, bandgap energy level, optical and other traits (Khlyustova et al., 2020). For example, recent research shows that zinc oxide ZnO (wurtzite); has higher chemical stability, good electron activity at room temperature, non-toxicity, and has a broader bandgap energy (3.2 eV) compared to $TiO_2$ phases [11]. However, single usage of ZnO also has limitations, such as the recombination rate of electrons and holes difference and photosensitivity [12]. However, the coupling of semiconductor oxides provides enhanced redox reactions that create increased photocatalytic activity. In addition, coupling $TiO_2$ with ZnO tend to enhance chemical stability and change the recombination rate of electron and holes [13]. Furthermore, the difference in valence and conduction band of $TiO_2$ and ZnO generates a more stable conductivity [14]. In the antibacterial inhibition part, providing a redox reaction increases the antibacterial inhibition because of the ROS generations [15]. These features of ZnO doped $TiO_2$ thin films may provide enhanced optical, structural, and antibacterial activity. In the production of $TiO_2$, various methods can be helpful, but the most important thing is time and money consumption. Recently, most of the technological advancement done by the nano-thickness scale and these materials provide various biomedical, electrical, physical, and chemical abilities. These new properties can be beneficial for bioactivity. The introduction of thin films on bioactivity can be beneficial for various applications. A single or multilayer creates these films and their thickness changes from nanometers to a few micrometers. These thin films contain two parts: the substrate and the layer part. These thin films provide various functions to execute against microorganisms, such as material transformation on wastewater's harmful environment or biodegradation. Due to these unique functionalities, the preparation of thin films is crucial for the outcome. Different deposition techniques are viable in thin-film production, such as physical and chemical depositions[15-21].

In this study, the sol-gel method was chosen as a preparation technique since the sol-gel method is a beneficial deposition technique to obtain a thin film for several different usage areas such as biomedical, chemical and electrical[18] and its unique traits such as homogeneity, sustainability, effectiveness, and low cost [22].

Homogeneity is crucial for this experiment because this experiment mainly focused on comparison of ZnO doped $TiO_2$ thin films with pure $TiO_2$ thin films.

## 2. MATERIALS AND METHOD

### 2.1. Production of Brookite Phase

The nanostructured $TiO_2$ sol was prepared using a mixture of titanium tetra isopropoxide (Ti $[OCH(CH_3)_2]_4$; (TTIP) Sigma-Aldrich), nitric acid ($HNO_3$), deionized water (DI), and isopropanol. First, titanium tetra isopropoxide (TTIP) was dissolved in isopropanol. Nitric acid was added dropwise in the solution under continuous stirring. Deionized water was added for hydrolysis. As a solution of TTIP:isopropanol:DI:AcAc a volume ratio of (0.4:4:0.1:0.2) was used. The solution was mixed using magnetic stirring for three hours at a room temperature of 22 $^oC$.

### 2.2 Production of Anatase Phase

Titanium tetra isopropoxide (Ti $[OCH(CH_3)_2]_4$; (TTIP) was mixed with acetic acid and ethanol with a molar ratio of 0.08:0.48:4, respectively. The mixture was stirred for two hours at a room temperature of 22 $^oC$.

### 2.3 Production of Rutile Phase

Tetrabutyl orthotitanate (Ti $(OCH_2CH_2CH_2CH_3)_4$; TBOT) was dissolved in ethanol under vigorous stirring for 30 minutes. Then, in another beaker, ethanol was mixed with Hydrochloric acid (HCl) for 30 minutes. Afterward, the second solution in the other beaker was slowly added dropwise to the first solution under vigorous stirring until homogeneous. Then, preparation of the $TiO_2$ films with various crystal phases, the solutions prepared by the sol-gel process were spin-coated on the corning 2947 substrates. Next, the films were heat-treated at 450 $^oC$ for one hour, for brookite and anatase crystal phases, at 600 $^oC$ for the rutile crystal phase. After the heat treatment, one layer was formed on the surface of the films. Then, the films were coated with a spin coater again, paying attention to the $TiO_2$ solution coated side on the upper surface, and heat treatment and spin coating processes were continued until the films were 3-layered.

### 2.4 Production of ZnO (wurtzite)

The solution was prepared by dissolving zinc acetate dehydrate (ZnAc) in isopropanol. Dea (Diethanolamine), which is a surface-active material, is used to accelerate solving. Water was added for hydrolysis reactions, as a precursor solution of ZnAc:isopropanol:Dea: water, a volume ratio of 0.4:4:0.1:0.2 was used. The solution was mixed using magnetic stirring for one hour at 60°C.

### 2.5 Production of ZnO/$TiO_2$ material

The obtained solution in Section 3.1.4 was mixed with one of the leading $TiO_2$ solutions for brookite, rutile, or anatase phase at ZnO/$TiO_2$ volume ratios of 0, 0.01, 0.02, 0.05, 0.1, denoted as ZTA0-10, ZTB0-ZTB10, ZTR0-ZTR10, respectively. The final solutions were deposited on corning 2947 glass substrates by spin-coating deposition (1000 rpm/30 s), using a spin coater at room temperature (22$^o$C). After coating, ZnO films were immediately placed in a microprocessor controlled (CWF 1100) furnace, heated at 450$^o$C. The films were taken out of the furnace and left at room temperature at the end of 1 hour. Finally, all coatings and heat treatment processes were repeated two times to get three-layered films.

### 2.6 Optical Properties and Homogeneity Studies

Ultraviolet-visible (UV-vis) spectrophotometer is an absorbance determination material used to understand materials' optical features in solution or solid phase. This device tracks the excited atoms by looking at their wavelength and absorbance energies. UV-vis spectrophotometer detects the absorbance of the film between the range of 200-1100 nm spectrum. The absorbance value is calculated by Beer-Lambert law (Equation 1).

$$A = \log_{10} (I_0/I) = \varepsilon bc, \qquad (1)$$

For the single wavelength, A is absorbance (unitless), ε is known as a molar absorptivity (molar, $M^{-1}cm^{-1}$), b is the length of the path (cm), c is solution concentration (M), $I_0$ is the intensity of light at a specific wavelength, and $I$ is the transmitted intensity. Moreover, UV-vis spectroscopy can provide quantitative and qualitative information about samples.

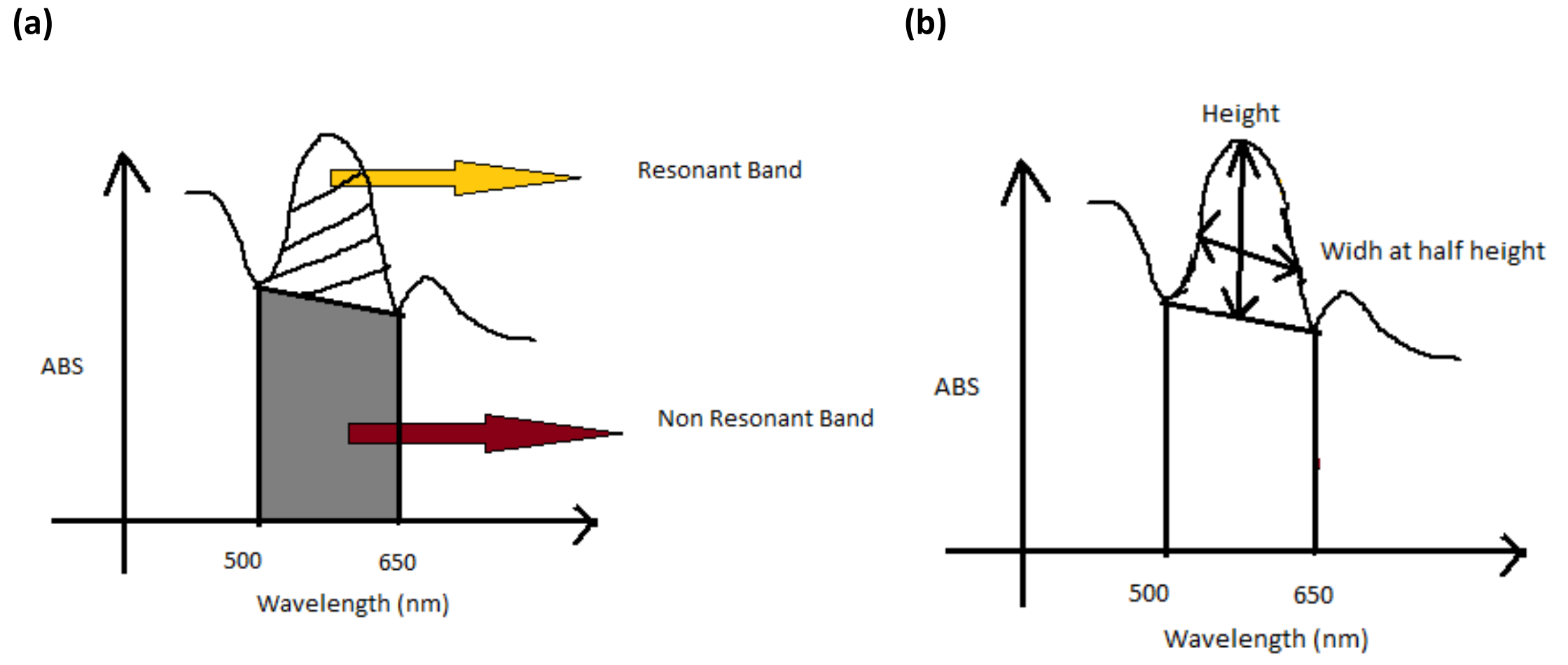

**Figure 1. (a)** Representation of resonant and non-resonant band **(b)** Representation of Height and Width at half height (FWHM)

In UV-vis spectrophotometric analysis' the most crucial part is finding the area under the curve because the area under the curves gives the resonance ratio. The resonance ratio is an essential concept for homogeneity, and it has been calculated by dividing the resonant band area into a non-resonant band area. In addition, another critical concept called normalized width is calculated by the division of width of the resonant band into the height of the resonant band-both resonance ratio and normalized width is essential for optical analysis of materials [30].

Resonance ratio = Area of resonant band/ area of the nonresonant band (2)

Normalized width= Width of resonance band/ height of the resonant band (3)

## 2.7 XRD Analysis

XRD is a widely used analytic method for physical and chemical composition identification. XRD instruments include three elements: a sample holder, an X-ray detector, and an X-ray tube. Generally, cathode tubes create the X-ray and voltage applied for accelerating the electrons than electrons bombarding the target material. Therefore, the XRD is a helpful method for analyzing homogeneity, crystal property, and average particle size. In general, XRD creates monochromatic X-rays, and these X-rays are scattered at specific angles to create specific peaks according to a sample's crystalline structure 31.

Every material has a unique set of d-spacing which means converting diffraction peaks to d-spacing reveals the type of the material. Due to these traits, the XRD pattern reveals the atomic composition of the material. Moreover, XRD analysis provides the atomic position of the crystalline structure. XRD analysis application on thin films provides information about lattice formation between substrate and film, homogeneity of materials, quality of the film, thickness, and toughness of the thin film. The XRD result shows the average crystal size using Scherrer’s equation (Equation 4).

The particle diameter size of different thin films was calculated according to Equation 4:

$$D = \frac{K\,\lambda}{B\cos\theta} \tag{4}$$

In Equation 4, D represents the diameter of the $TiO_2$ nanocrystals, K presents the shape factor, and it has a constant value (0.89), λ represents the radiation wavelength, B represents the full width at half maximum (FWHM) of the diffraction line, and θ (theta) represents the Bragg angle (rad) [32].

## 2.8 Antibacterial Measurement

The antibacterial activity of $TiO_2$ is crucial for biodegradation, wastewater remediation and antibacterial disinfection. $TiO_2$ thin films used various applications related to antibacterial activity. The advantages of $TiO_2$ are photocatalytic activity, high degradation efficiency and non-toxicity [33]. Due to these traits, $TiO_2$ thin film can provide a suitable solution for antibacterial disinfection studies. Antibacterial materials have rapidly grown in recent years, and most technological advancements need to provide antibacterial traits. Therefore, the ISO 22196 method was used to determine the antibacterial activity of $TiO_2$ thin films. In addition, ISO 22196 is a method that quantitatively evaluates the antibacterial activity (growth inhibition or killing) of samples such as plastic, other non-porous, and surface products (ICS, 2011). Therefore, ISO 22196 is an excellent method to determine bacteriostatic (growth inhibition) and bactericidal (killing bacteria) features, and this test is highly critical for $TiO_2$ usages for future technologies. Different types of pure and doped $TiO_2$ thin film antibacterial efficiencies were represented (Table 1).

**Table 1.** Antibacterial efficiency of different types of Pure $TiO_2$ and doped $TiO_2$

| Thin Film type | Bacteria type | Light Source | Highest Antibacterial inhibition % | References |
|---|---|---|---|---|
| $Sn^{+4}$ - $TiO_2$ film | *E. coli, S.aureus* | UV | 99.9% | 34 |
| $TiO_2$ film | *S.aureus, S.epidermidis, E.coli* | UV | 47 % | 35 |
| $TiO_2$ /Ag/Cu | *E. coli* | UV | 100% | 36 |
| Mg- $TiO_2$ thin film | *E. coli Pseudomonas Bacillus sp Staphylococcus* | UV | 100% | 37 |
| Cu-doped $TiO_2$ thin film | *Phytophthorapalmivorahas* | UV | 75% but concentrations >3% kill %100 | 38 |
| Ag- $TiO_2$ /PDMS thin film | *M.luteus S.maltophilia* | Visible irradiation | 100% | 39 |
| ZnO(wurtzite), $TiO_2$ , and ZnO- $TiO_2$ thin films | *A.flavusthan* | UV | 100% | 40 |

Surfaces needed to be smooth,1cm thick, and 5x5 $cm^2$ shapes. Six samples of a corning glass covered with thin films were prepared for the experiment; three samples were to test bacteria and three to control groups. Each of these samples ($TiO_2$, ZnO/$TiO_2$) was analyzed by ISO 22196.

In antibacterial analysis, gram-positive "*Staphylococcus aureus*" and gram-negative "*Escherichia coli*" were used. The antibacterial activity of the films was evaluated by colony-forming unit (CFU) counting. After incubation, the colonies were counted. CFU per ml was calculated for each sample at different time intervals (0-120 min) by using the following formula:

$$\text{CFU/ml} = \text{No. of colonies} \times \text{Dilution factor} / \text{volume inoculated} \quad (5)$$

### 2.9 Characterizations

The absorbance spectra of composite films were measured using Labomed Spectro 22 UV–vis Spectrophotometer in the spectral range of 190–1100 nm wavelengths. The calculations of under the curve area, the entire width of half maximum of the absorbance data were performed by Origin 8.0 Software focusing the absorbance peak regions. The optical band gap energies of the composite films were calculated from the Tauc Plot since the thickness of the films was suitable for applying Beer-Lambert's Law. An X-ray diffractometer (XRD, Philips PW-1800) with Cu-Kα radiation (The wavelength of Cu-Kα is 0.15406 nm) was used to identify the films' crystal phases. The antibacterial response of the films was investigated using gram-positive "*Staphylococcus aureus*" (*S. aureus*) and gram-negative "*Escherichia coli*" (*E. coli*) according to the standard of ISO 22196. The initial dose of these bacteria's concentration was $10^5$CFU/ml. Since $TiO_2$ shows antibacterial properties under UV light exposure and its photocatalytic effect is known, the antibacterial effect in the visible light region was investigated by doping with ZnO.

## 3. RESULTS AND DISCUSSION

### 3.1 XRD Results

The most crucial thing in the calculation is the theta value. Generally, theta value is found with two thetas and must convert a single theta value to determine the diameter of particles properly. Then, the calculation and comparison of the XRD values were determined (Table 2).

**Table 2. XRD values of different types of thin films and their diameter size**

| Thin films | K | λ(A) | Peak positions 2θ | FWHM | D(nm) |
|---|---|---|---|---|---|
| $TiO_2$ (anatase) | 0.89 | 1.54 | 25.27 | 0.36 | 21.85 |
| $TiO_2$ (brookite) | 0.89 | 1.54 | 31.58 | 0.16 | 50.18 |
| $TiO_2$ (rutile) | 0.89 | 1.54 | 27.30 | 0.21 | 36.89 |
| ZTA0(1$^{st}$ Peak) | 0.89 | 1.54 | 25.27 | 0.36 | 21.85 |
| ZTA0(2$^{nd}$ Peak) | 0.89 | 1.54 | 37.69 | 0.34 | 24.19 |
| ZTA1(1$^{st}$ Peak) | 0.89 | 1.54 | 25.23 | 0.31 | 25.87 |
| **ZTA1(2nd Peak)** | 0.89 | 1.54 | 37.66 | 0.25 | 32.05 |
| **ZTA2 (1st Peak)** | 0.89 | 1.54 | 25.21 | 0.36 | 21.77 |
| **ZTA2 (2nd Peak)** | 0.89 | 1.54 | 37.67 | 0.25 | 32.12 |
| **ZTA5** | 0.89 | 1.54 | 25.19 | 0.43 | 18.66 |
| **ZTA10(1st Peak)** | 0.89 | 1.54 | 25.22 | 0.34 | 23.45 |
| **ZTA10(2nd Peak)** | 0.89 | 1.54 | 37.68 | 0.33 | 24.75 |
| **ZTA10(1st Peak** | 0.89 | 1.54 | 25.23 | 0.31 | 25.87 |
| **ZTA10(2nd Peak)** | 0.89 | 1.54 | 37.66 | 0.25 | 32.05 |
| **ZTA10(3rd Peak)** | 0.89 | 1.54 | 25.21 | 0.36 | 21.77 |
| **ZTA10(4th Peak)** | 0.89 | 1.54 | 37.67 | 0.25 | 32.12 |
| **ZTB0** | 0.89 | 1.54 | 31.58 | 0.16 | 50.18 |
| **ZTB1** | 0.89 | 1.54 | 31.54 | 0.35 | 23.07 |
| **ZTB2** | 0.89 | 1.54 | 31.55 | 0.34 | 23.97 |
| **ZTB5** | 0.89 | 1.54 | 31.60 | 0.48 | 16.98 |
| **ZTB10** | 0.89 | 1.54 | 31.76 | 0.59 | 13.61 |
| **ZTR0** | 0.89 | 1.54 | 27.30 | 0.21 | 36.89 |
| **ZTR1(1st Peak)** | 0.89 | 1.54 | 27.30 | 0.24 | 33.22 |
| **ZTR1(2nd Peak)** | 0.89 | 1.54 | 36.18 | 0.55 | 14.93 |
| **ZTR2(1st Peak)** | 0.89 | 1.54 | 27.30 | 0.21 | 36.96 |
| **ZTR2(2nd Peak)** | 0.89 | 1.54 | 36.08 | 0.25 | 32.86 |
| **ZTR5** | 0.89 | 1.54 | 27.30 | 0.21 | 37.21 |
| **ZTR10(1st peak)** | 0.89 | 1.54 | 36.08 | 0.28 | 29.49 |
| **ZTR10(2nd Peak)** | 0.89 | 1.54 | 25.76 | 5.58 | 1.44 |
| **ZTR10(3rd Peak)** | 0.89 | 1.54 | 31.59 | 0.92 | 8.85 |
| **ZTR10(4th Peak)** | 0.89 | 1.54 | 34.40 | 0.62 | 13.08 |
| **ZTR10(5th Peak)** | 0.89 | 1.54 | 36.19 | 0.64 | 12.73 |
| **ZnO(1st Peak)** | 0.89 | 1.54 | 31.75 | 0.15 | 52.87 |
| **ZnO((2nd Peak)** | 0.89 | 1.54 | 34.55 | 0.16 | 50.56 |
| **ZnO(3rd Peak)** | 0.89 | 1.54 | 36.44 | 0.16 | 51.11 |

In this study, decreased size diameter needs to be achieved after using ZnO as a dopant material. The average particle size of ZnO was found at 51.51 nanometers. The average size of pure $TiO_2$ phases anatase was at 21.86 nm, brookite was at 50.18 nm, and rutile was at 36.9 nm. ZnO doped anatase TiO2 (ZTA) XRD data shows that a small concentration of ZnO doped increases the diameter size of $TiO_2$ then decreases the composite size slowly after adding more ZnO concentration. In ZnO doped brookie $TiO_2$ (ZTB), XRD data shows that adding every ZnO into brookite decreases the diameter size due to brookite orthomorphic structure. ZnO doped rutile TiO2 XRD data shows that the addition of ZnO also decreases the average particle size of pure slowly. This slow increase and decrease of particle size in ZTR (doped rutile) and ZTA (doped anatase) is related to their tetragonal structure. Symmetrical structure affects the bond formation and hindering.

Figure 2 (a) shows that the films have a rare brookite phase, an orthorhombic crystal structure. A single crystallization peak (211) of the brookite phase at a 2θ value of 31.5° corresponds to ICDD Card No. 04-019-9878. In Figure 2 (c), the sharp diffraction peaks at 25.2°, 37.7°, and 47.8° theta values are indexed to the (101), (004), and (200). Planes of the anatase phase of TiO2 correspond to 21-1272 according to ICDD Card. In Figure 2 (d), the peaks at the 2θ values of 27.31°, 36.12°, and 41.25° are fitted well with the (110), (101), and (111) planes of the rutile phase of $TiO_2$ (ICDD Card No. 21-1276), respectively. These (anatase and rutile) are tetragonal phases of $TiO_2$.

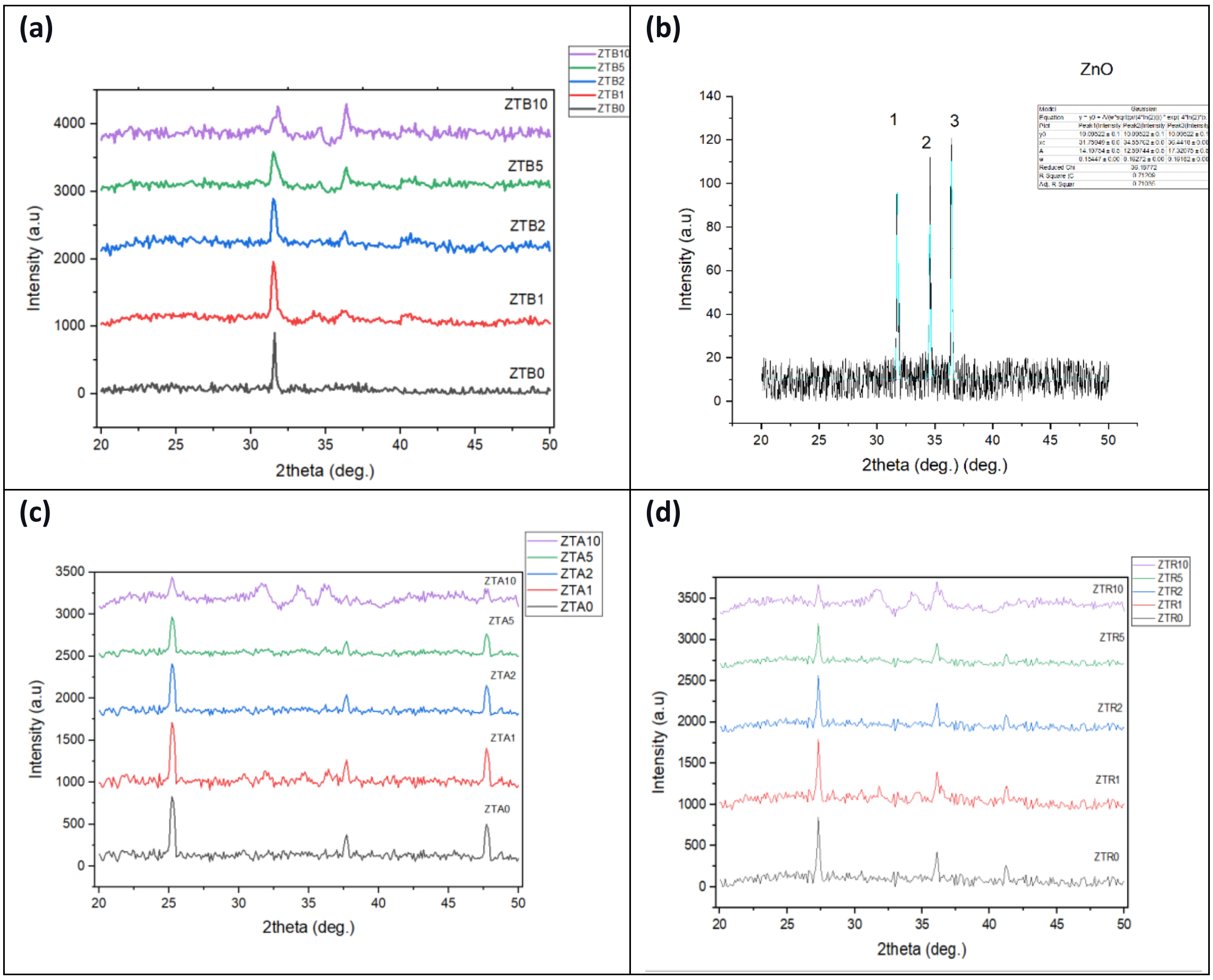

**Figure 2.** XRD response of **(a)** ZnO doped $TiO_2$ brookite films with various $ZnO/TiO_2$ doping ratios, **(b)** ZnO film and parameter calculations, **(c)** ZnO doped $TiO_2$ anatase films with various $ZnO/TiO_2$ doping ratios, **(d)** ZnO doped $TiO_2$ rutile films with various $ZnO/TiO_2$ doping ratios.

### 3.2 Optical Characteristics of the Films

Absorption spectra of pure $TiO_2$ films and ZnO doped $TiO_2$ films were analyzed using UV-vis spectrophotometers, as shown in Figure 3. High absorbance values are seen between at 200-500 range. According to absorbance-wavelength graphs, the addition of ZnO can enhance the absorbance rate of all three phases. On the other hand, the high absorbance value of ZnO can decrease the absorbance range of Pure $TiO_2$ phases because dilution of pure TiO2 absorbance ability reduces, and the addition of ZnO introduces its photon absorbance rate. Recombination of rate increases, and absorption of doped nanocomposites decreases. The addition of ZnO also affects the resonance ratio and homogeneity of the material. Doped material may enhance the photocatalytic activity, but the optimum amount of dopant addition is so important for homogeneity concerns. A comparison of different volumes of ZnO doped anatase, brookite, and rutile was determined (Figure 3 to 6). In normal conditions, the material homogeneity is determined by the relation between resonance ratio and normalized width. When the resonance ratio increases and normalized width decreases, that implies composites reach a more homogenous structure. According to calculations, adding ZnO dopant decreases the resonance ratio and increases the normalized width, which is vice versa of literature information. However, our studies do not ultimately increase or decrease; some ranges in which composites resonance ratio increases with ZnO. This result provides good information, such as the critical point to cut down dopant addition to getting the more homogenous structure. Further studies can arrange their dopant addition to these ranges to get more homogenous material. ZTA5-10, ZTB5-10, and ZTR5-10 concentration ranges are critical for the homogeneity shift for excessive dopant addition (Table 3).

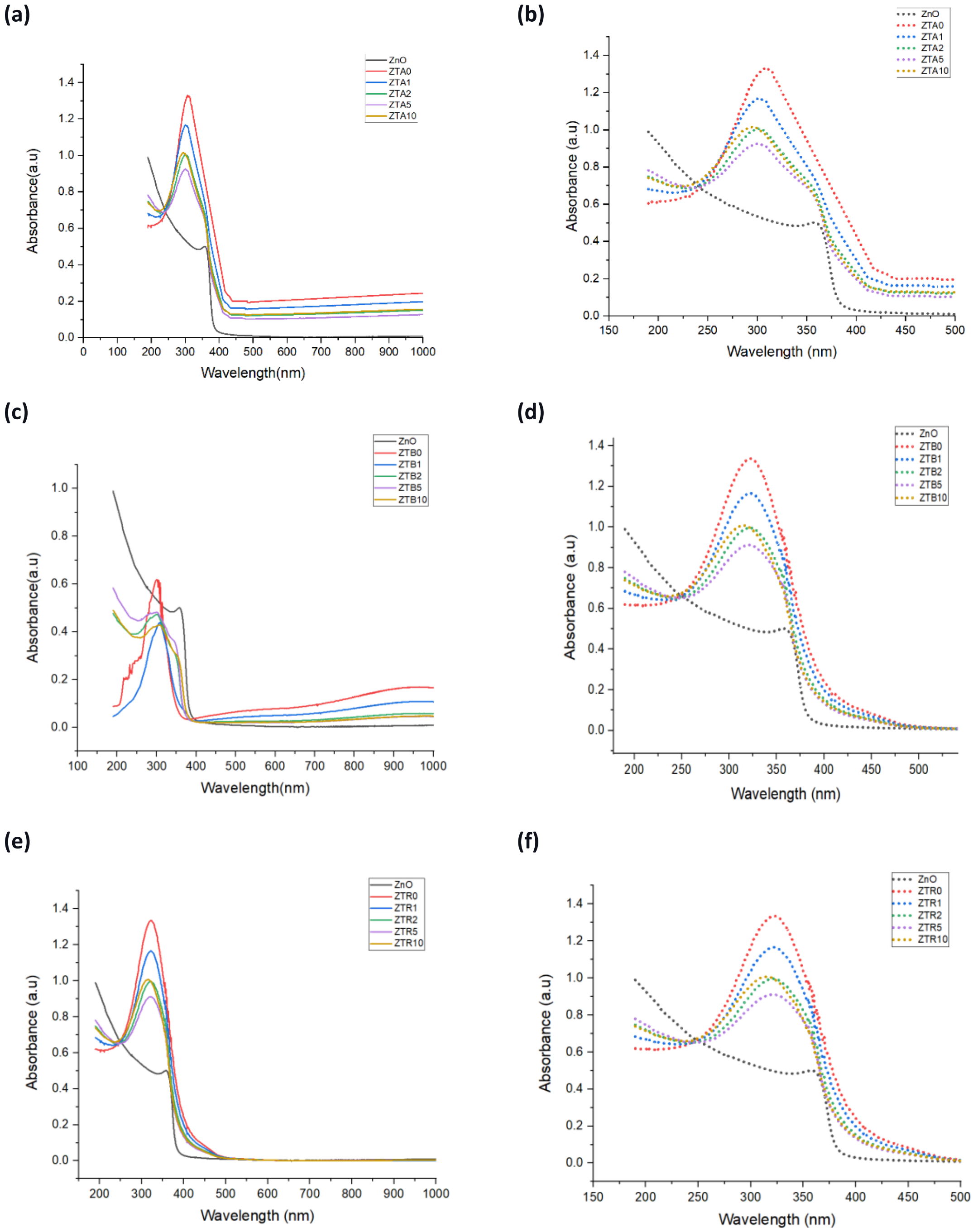

**Figure 3.** Absorbance wavelength spectrum of ZnO and ZnO doped $TiO_2$. In **(a-c-e)** graphs show the range between 200-1000 spectrum, and **(b-d-f)** graphs show the detailed range between 200-500 range spectrum.

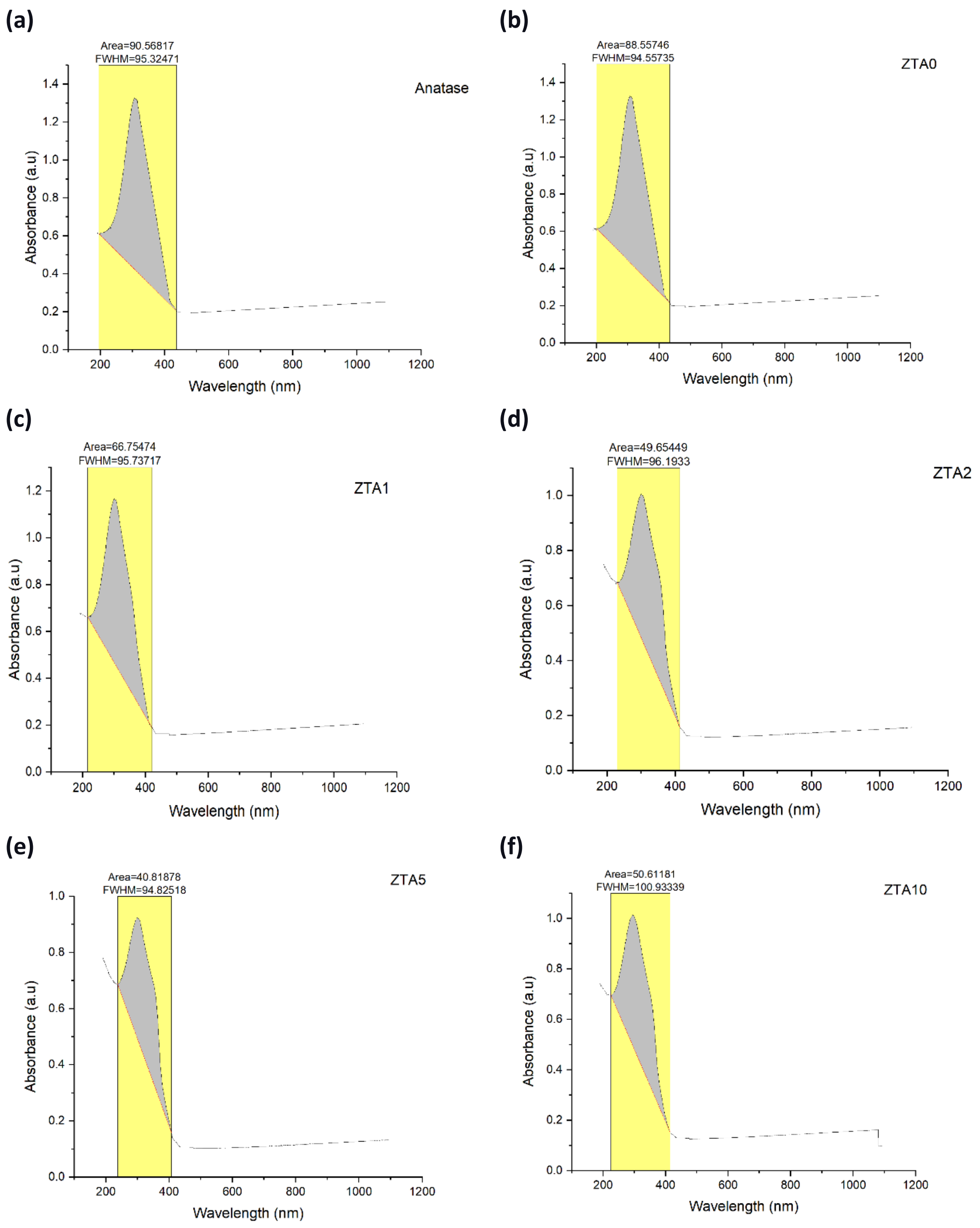

**Figure 4.** Resonant band representation of pure **(a)** and ZnO doped **(b-f)** anatase $TiO_2$

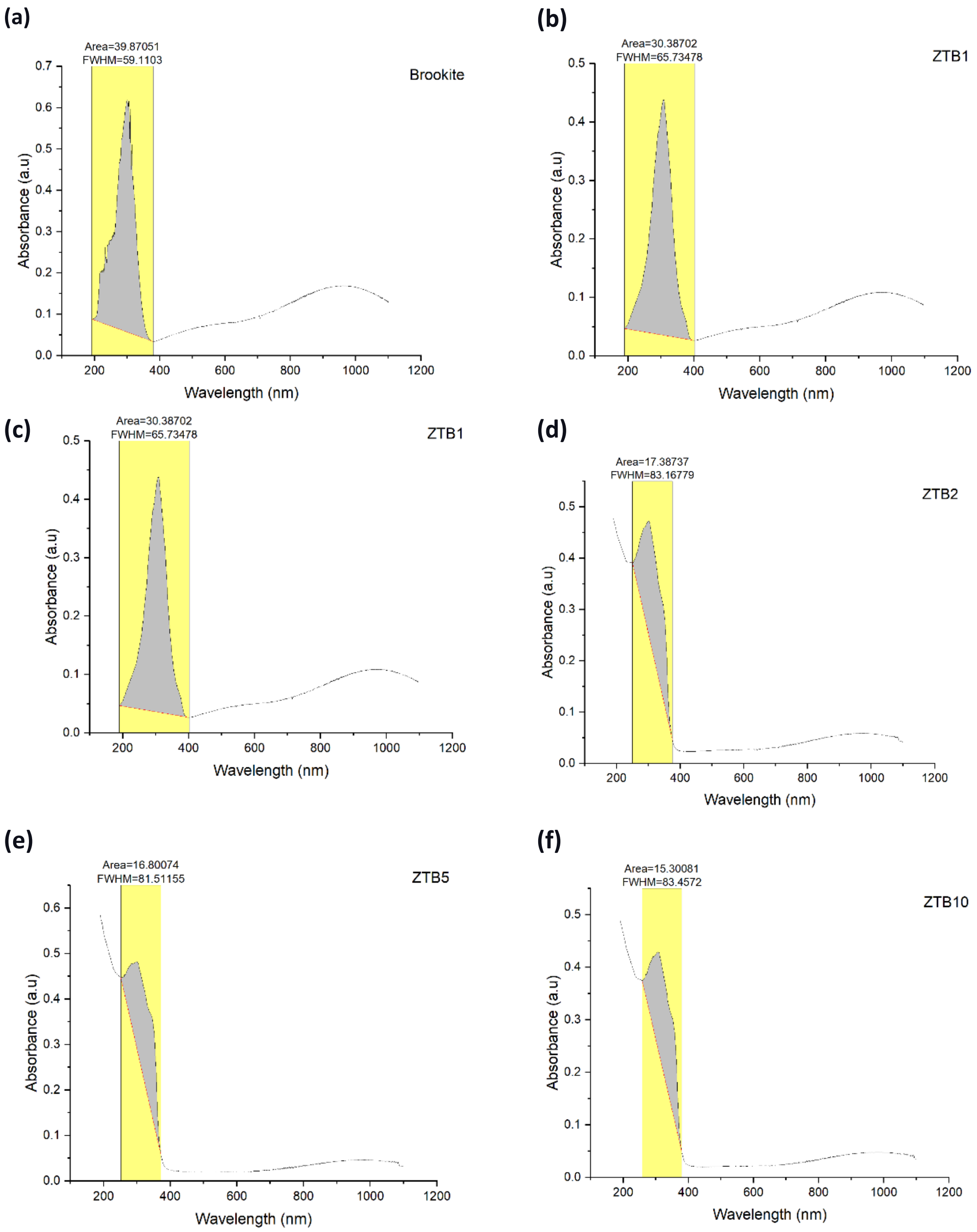

**Figure 5.** Resonant band representation of pure **(a)** and ZnO doped **(b-f)** brookite $TiO_2$

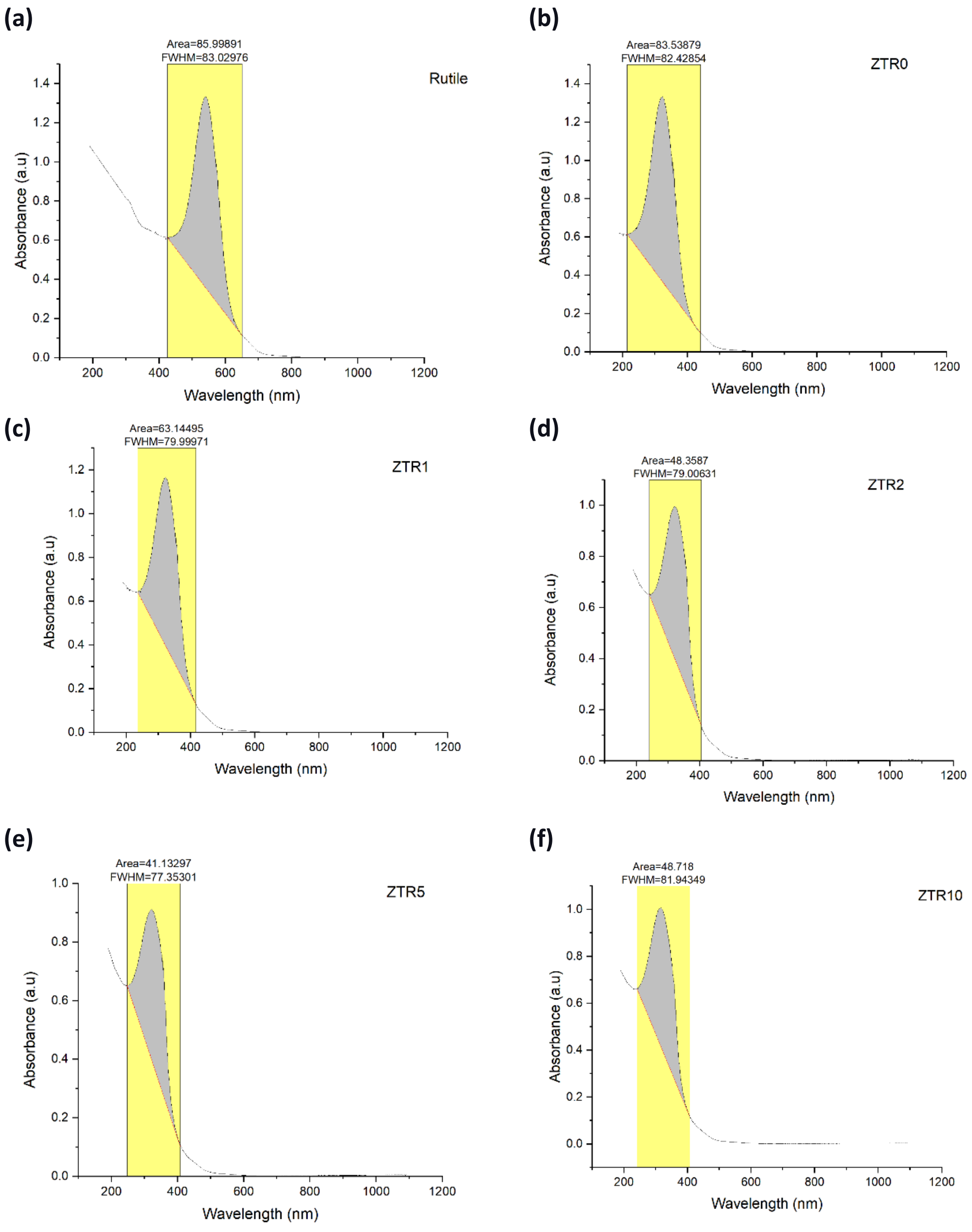

**Figure 6.** Resonant bands representation of pure **(a)** and ZnO doped **(b-f)** rutile $TiO_2$.

In-band energy analysis, the Tauc pilot measured materials' band-gap energies (Equation 6).

$$(\alpha hv) = B\ (hv\text{-}E_g)^r \qquad (6)$$

**Table 3.** Calculation of resonance ratio and normalized width of pure and ZnO doped $TiO_2$

| Thin films | Area of resonant band | Area of non-resonant background | Resonance ratio | Width of resonant band | Height of resonant band | Normalized width |
|---|---|---|---|---|---|---|
| **Pure $TiO_2$ (anatase)** | 90.56 | 100.52 | 0.90 | 95.32 | 0.90 | 104.97 |
| **Pure $TiO_2$ (brookite)** | 39.87 | 11.46 | 3.47 | 59.11 | 0.56 | 105.15 |
| **Pure $TiO_2$ (rutile)** | 85.99 | 87.29 | 0.98 | 83.02 | 0.98 | 84.11 |
| **ZTA0** | 88.55 | 97.13 | 0.91 | 94.55 | 0.89 | 105.14 |
| **ZTA1** | 66.75 | 87.41 | 0.76 | 95.73 | 0.70 | 135.66 |
| **ZTA2** | 49.65 | 77.41 | 0.64 | 96.19 | 0.53 | 180.24 |
| **ZTA5** | 40.81 | 72.15 | 0.56 | 94.82 | 0.44 | 212.55 |
| **ZTA10** | 50.61 | 80.79 | 0.62 | 100.93 | 0.52 | 191.56 |
| **ZTB0** | 39.90 | 11.34 | 3.51 | 59.13 | 0.56 | 105.20 |
| **ZTB1** | 30.38 | 7.75 | 3.91 | 65.73 | 0.40 | 163.06 |
| **ZTB2** | 17.38 | 27.52 | 0.63 | 83.16 | 0.22 | 364.62 |
| **ZTB5** | 16.80 | 30.29 | 0.55 | 81.51 | 0.21 | 378.41 |
| **ZTB10** | 15.30 | 35.76 | 0.42 | 83.45 | 0.19 | 431.66 |
| **ZTR0** | 83.53 | 81.09 | 1.03 | 82.42 | 0.96 | 85.17 |
| **ZTR1** | 63.14 | 70.08 | 0.90 | 79.99 | 0.77 | 103.60 |
| **ZTR2** | 48.35 | 64.85 | 0.74 | 79.00 | 0.60 | 130.06 |
| **ZTR5** | 41.13 | 60.50 | 0.67 | 77.35 | 0.52 | 146.87 |
| **ZTR10** | 48.71 | 64.83 | 0.75 | 81.94 | 0.60 | 136.11 |

In general, Titanium dioxide ($TiO_2$) is an excellent semiconductor in terms of bandgap energy, such as anatase 3.20 eV, rutile 3.00 eV, and brookite 3.13 eV [4-6]. The photocatalytic ability of pure $TiO_2$ and ZnO doped $TiO_2$ was determined. In a normal condition, ZnO addition needs to increase the bandgap energy of pure $TiO_2$, but $TiO_2$ is an indirect semiconductor, and ZnO is a direct conductor. Thus, every possible indirect and direct graph needs to be evaluated for optimum results (Figure 7). These graphs find from the absorbance value of every single material. In direct conduction, the r variable equals two, and the indirect conduction r variable equals to ½. After creating their Tauc plots, ZnO doped anatase tends to be a direct conductor, ZnO doped brookite tends to the indirect conductor, and ZnO doped rutile tend to act like a direct conductor. Discrimination of indirect and direct conduction is done by using the Tauc plot. ZnO normal bandgap energy around at 3.37 eV [11]. The addition of ZnO dopant needs to increase the original bandgap energy of pure $TiO_2$ phases. Due to this reason, (a), (d), (e) in Figure 7 shows that the addition of ZnO dopant increases the original band gap energy of $TiO_2$.

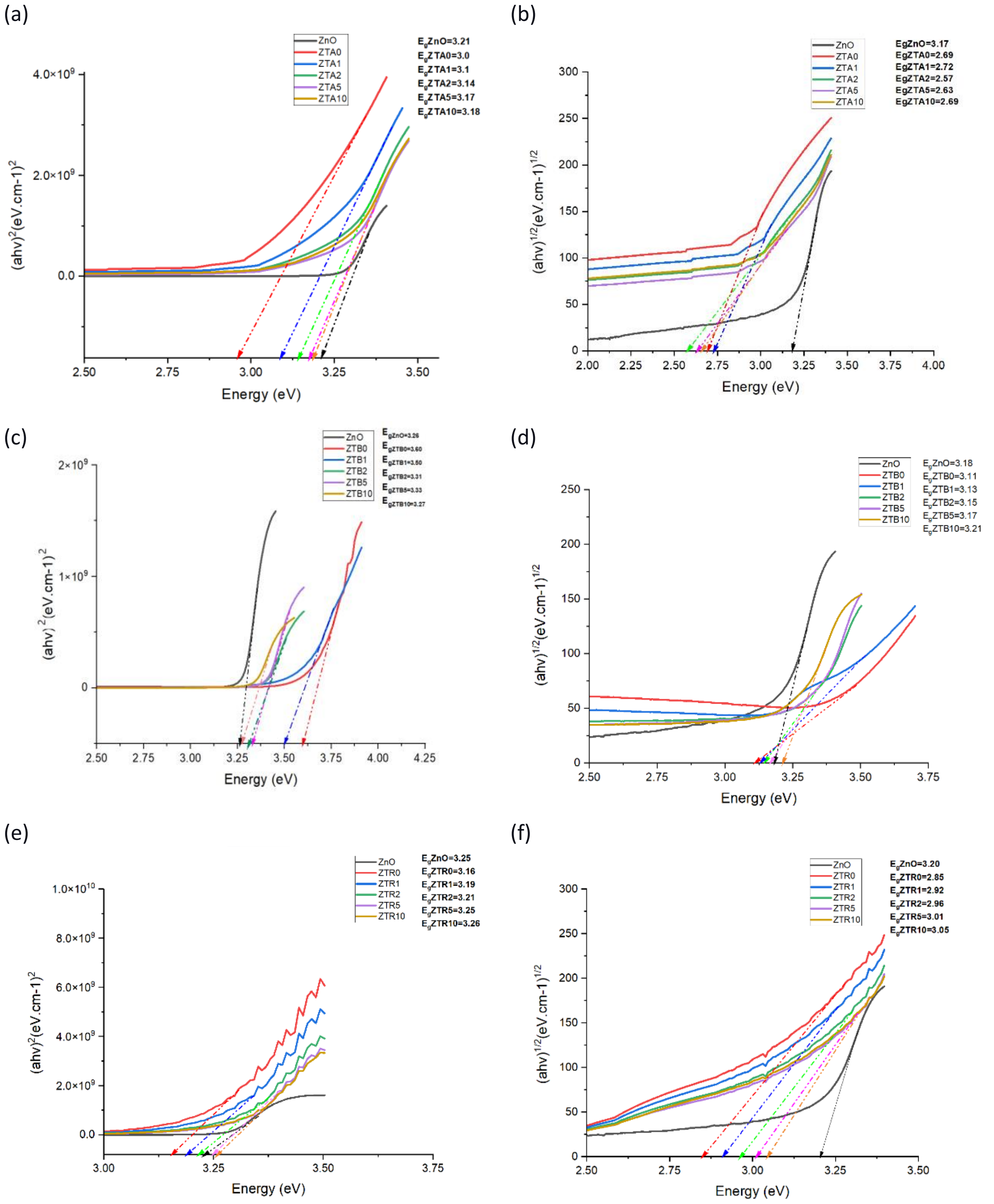

**Figure 7.** Tauc Plot of pure ZnO and ZnO doped Anatase, Brookite, and Rutile phases of $TiO_2$ with various $ZnO/TiO_2$ ratios. The extrapolation of the curves shows the optical energy band gap of ZTA, ZTR, and ZTB.

**Table 4.** Bandgap energy values of different types of doped material derived from the Tauc plot.

| Film type | Direct(eV) | Indirect(eV) |
|---|---|---|
| **ZnO** | 3.21 | 3.17 |
| **ZTA0** | 3.0 | 2.69 |
| **ZTA1** | 3.1 | 2.72 |
| **ZTA2** | 3.14 | 2.57 |
| **ZTA5** | 3.17 | 2.63 |
| **ZTA10** | 3.18 | 2.69 |
| **ZTB0** | 3.60 | 3.11 |
| **ZTB1** | 3.50 | 3.13 |
| **ZTB2** | 3.31 | 3.15 |
| **ZTB5** | 3.33 | 3.17 |
| **ZTB10** | 3.27 | 3.21 |
| **ZTR0** | 3.16 | 2.85 |
| **ZTR1** | 3.19 | 2.92 |
| **ZTR2** | 3.21 | 2.96 |
| **ZTR5** | 3.25 | 3.01 |
| **ZTR10** | 3.26 | 3.05 |

### 3.3 Antibacterial Activity

Figure 8 presents the antibacterial response of films with respect to the time for *E. coli* and *S.aureus*. *E. coli* count reduced from $10^5$ to $10^4$ with 120 min of irradiation time for pure brookite $TiO_2$ film. A dramatic decrease in the bacteria number for both bacteria was observed for ZnO/TiO2 composite films compared with $TiO_2$ films under the lighting with visible light. After 120 min. of light exposure, the bactericidal survival decreases of ZTA0, ZTA1, ZTA2, ZTA5, and ZTA10 films against E. coli were $10^0$, $3x10^1$, $10^2$, $2x10^2$ and $7x10^3$, respectively. Meanwhile, exhibiting a similar trend, the bactericidal survival decreases of ZTA0, ZTA1, ZTA2, ZTA5, and ZTA10 films against S. aureus were $10^0$, $5x10^1$, $2x10^2$, $6x10^3$ and $10^4$, respectively. For ZTB2-ZTB10 films, complete inactivation of the *E. coli* was recorded at 120, 90, and 75 min of light exposure, respectively. Since $TiO_2$ shows antibacterial properties under UV light exposure and its photocatalytic effect is known, the antibacterial effect in the visible light region was investigated by doping with ZnO. These results show that the content of ZnO in the ZnO/$TiO_2$ composite films plays a significant role for enhancing the antibacterial activity of the films in the absence of UV light.

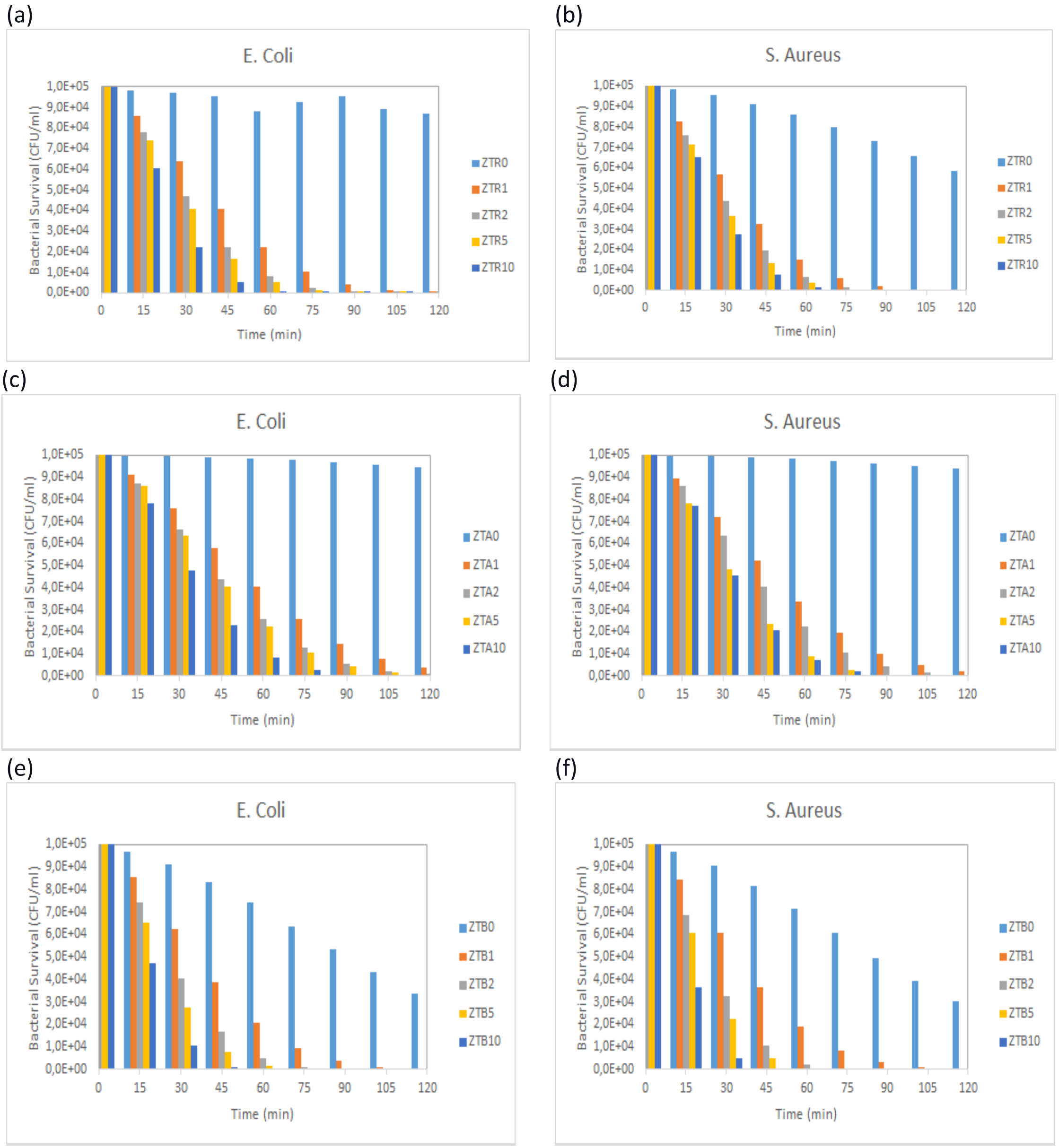

**Figure 8.** Antibacterial response of ZnO doped Anatase, Brookite, and Rutile phases of $TiO_2$ with various $ZnO/TiO_2$ ratios under light exposure for *E. Coli* and *S. Aureus*.

## 4. CONCLUSION

In conclusion, XRD results show that the particle size of $TiO_2$ can be reduced in the brookite phase rather than anatase and rutile due to brookite's large volume. However, anatase and rutile also show remarkable diameter size drop when the concentration is appropriately arranged. However, the ZTA5 sample is also a good concentration for anatase applications. Due to their structural stability traits, anatase and rutile particle size decrease slowly. UV data show that the resonance ratio is essential for cell homogeneity and dispersibility. Due to this reason addition of dopant materials needs to be controlled by stoichiometry for optimum homogeneity. Using the sol-gel method provides the minor homogeneity range for a concentration of doped ZTA5-10, ZTB5-10, and ZTR5-10. In these concentration ranges, particles size dropping and their antibacterial activity increases. Band gap energy enhancement of ZnO was shown by using the Tauc plot. In addition, direct ZnO doped anatase $TiO_2$, direct ZnO doped rutile and indirect ZnO doped brookite conduction bands were determined. In antibacterial inhibition, after 105 minutes, all of the bacterial growth inhibited against *E.coli*. After 90 minutes, all of the bacterial growth inhibited against *S.Aureus* in the rutile phase. After 120 minutes, most bacterial growth inhibited beside ZTA1 against *E.coli* and *S.Aureus* in the anatase phase. Finally, after 75 minutes, most bacterial growth inhibited beside ZTB1 against *E.coli* and *S.Aureus* in the brookite phase. This research mainly focused on discovering how dopant material can enhance the activity of pure material and how these enhanced traits can be used on optical and antibacterial activity. Therefore, ZnO/$TiO_2$ composite films can give an alternative solution for antibacterial surface applications under visible light.